\documentclass[12pt,preprint]{aastex}

\usepackage{epstopdf, graphicx, natbib}

\renewcommand\refname{References and Notes}
\bibpunct{(}{)}{,}{a}{}{;}

\def\lya{\ifmmode {\rm Ly}\alpha~ \else Ly$\alpha$~\fi}
\def\lyan{\ifmmode {\rm Ly}\alpha \else Ly$\alpha$\fi}
\def\lyb{\ifmmode {\rm Ly}\beta~ \else Ly$\beta$~\fi}
\def\lyg{\ifmmode {\rm Ly}\gamma~ \else Ly$\gamma$~\fi}
\def\arxiv{{{\rm Ar}\,{\sc xiv}~}}

\def\caxvii{{{\rm Ca}\,{\sc xvii}~}}
\def\caxvi{{{\rm Ca}\,{\sc xvi}~}}
\def\civ{\ifmmode {\rm C}\,{\sc iv}~ \else C\,{\sc iv}~\fi}
\def\civn{\ifmmode {\rm C}\,{\sc iv}~ \else C\,{\sc iv}\fi}

\def\cvin{\ifmmode {\rm C}\,{\sc vi} \else C\,{\sc vi}\fi}

\def\ovi{{{\rm O}\,{\sc vi}~}}
\def\ov{{{\rm O}\,{\sc v}~}}

\def\ovin{{{\rm O}\,{\sc vi}}}
\def\ovii{{{\rm O}\,{\sc vii}~}}
\def\oviii{{{\rm O}\,{\sc viii}~}}
\def\ovin{{{\rm O}\,{\sc vi}}~}
\def\oviin{{{\rm O}\,{\sc vii}}}

\def\fevii{{{\rm Fe}\,{\sc vii}~}}

\def\fevii-xii{{{\rm Fe}\,{\sc vii-xii}~}}

\def\fexvii{{{\rm Fe}\,{\sc xvii}~}}

\def\fexxv{{{\rm Fe}\,{\sc xxv}~}}

\def\fexxvi{{{\rm Fe}\,{\sc xxvi}~}}

\def\chandra{{\it Chandra}~}

\def\ark564{\object{Ark~564}}
\def\mrk590{{\it Mrk~590}}
\def\ngc3783{{\it NGC~3783}}

\title{Discovery of Relativistic Outflow in the Seyfert  
Galaxy \textbf{\textit{Ark 564} }}

\author{A. Gupta and S. Mathur\altaffilmark{1}}
\affil{Astronomy Department, Ohio State University, Columbus, OH 43210, USA}
\email{agupta@astronomy.ohio-state.edu}
\author{Y. Krongold}
\affil{Instituto de Astronomia, Universidad Nacional Autonoma 
de Mexico, Mexico City, (Mexico)}
\author{F. Nicastro}
\affil{Harvard-Smithsonian Center for Astrophysics, Cambridge, MA, 02138, USA}
\affil{Osservatorio Astronomico di Roma-INAF, 
Via di Frascati 33, 00040, Monte Porzio Catone, RM, (Italy)}

\altaffiltext{1}{Center for Cosmology and
Astro-Particle Physics, The Ohio State University, Columbus, OH 43210}

\begin{document}

\begin{abstract}
We present \chandra high energy transmission grating spectra of the
narrow-line Seyfert-1 galaxy Ark 564. The spectrum shows numerous
absorption lines which are well modeled with low velocity outflow
components usually observed in Seyfert galaxies \citep{Gupta2013}.
There are, however, some residual absorption lines which are not
accounted for by low-velocity outflows. Here we present identifications
of the strongest lines as $K\alpha$ transitions of \oviin (two lines)
and \ovin at outflow velocities of $\sim 0.1c$. These lines are detected
at $6.9\sigma$, $6.2\sigma$, and $4.7\sigma$ respectively and cannot be
due to chance statistical fluctuations. Photoionization models with
ultra-high velocity components improves the spectral fit significantly,
providing further support for the presence of relativistic outflow in
this source. Without knowing the location of the absorber, its mass and
energy outflow rates cannot be well constrained; we find 
$\dot{E}(outflow)/L_{bol}$ lower limit of $\geq 0.006$\% assuming a 
bi-conical wind geometry. 
This is the first time that absorption lines with
ultra-high velocities are unambiguously detected in the soft X-ray
band. The presence of outflows with relativistic velocities in AGNs with
Seyfert-type luminosities is hard to understand and provides valuable
constraints to models of AGN outflows. Radiation pressure is unlikely to
be the driving mechanism for such outflows and magneto-hydrodynamic may
be involved.
\end{abstract}

\section{Introduction}

There have been reports of ultra-high velocity outflows in the X-ray
spectra of radio-quiet AGNs and quasars. These outflows are identified
through blue-shifted \fexxv and/or \fexxvi absorption lines (at
rest-frame energies of 7-10 keV) from highly ionized gas ($\log
\xi\footnote{The ionization parameter $\xi=L_{ion}/nr^{2}$, where
$L_{ion}$ is the ionizing luminosity between 1 Ryd and 1000 Ryd (1
Ryd=13.6 eV), {\bf n} is the number density of the material and {\bf r}
is the distance of the gas from the central source.}=3 - 6~$
erg~s$^{-1}~$cm), with column densities as large as $N_{H} =
10^{22}-10^{24}~$cm$^{-2}$, and with relativistic velocities of
0.1c--0.3c \citep[and references therein]{Pounds2003a, Pounds2003b,
Tombesi2010}.  The mass outflow rate of these high velocity outflows
were comparable to the accretion rate and their kinetic energy was a
significant fraction of the bolometric luminosity
\citep{Tombesi2012}. These outflows can provide effective feedback that
is required by theoretical models of galaxy formation to solve a number
of astrophysical problems ranging from cluster cooling flows to
structures of galaxies.

Although high velocity outflows are detected in a large number of
sources, in all the cases identification are based on Fe K-shell
transitions which fall in a region of the spectrum where instrumental
resolution is much lower than in the soft X-ray band. The significance
of the absorption line detections is often questioned and with only a
few lines observed, accurate parametrization of the photoionized plasma
becomes difficult. While the observed transitions are from highly
ionized gas ($\log \xi = 3-6~$ers~s$^{-1}~$cm), photoionization models predict
that in such a plasma in addition to highly ionized iron, lighter
elements should also be present, strongest of which are S, Si and O
\citep[e.g.,][]{Sim2010, Krongold2003}. Absorption lines from these highly
ionized elements lie in the soft X-ray band, so warm absorber signatures
of relativistic outflows at soft X-rays should be present, but were
never unambiguously observed until now (see \S 5.3 for existing
evidence). For complete understanding of the properties of high velocity
outflows it is necessary to detect transitions of other ions such as of
O, Ne, Si and S as predicted by models.

In this paper, we report the serendipitous discovery of high velocity
outflows in the soft X-ray band in \ark564, identified during our
detailed analysis of the \chandra archival data of this source 
\citep[hereafter Paper~I]{Gupta2013}.

Primary goal of Paper~I was to self consistently
analyze and model the grating spectra of typical
($V_{out}=100-1000~$km~s$^{-1}$) warm absorbers in \ark564. Here we
present the discovery of warm absorbers (WA) with ultra-high velocities.

\section{Data and Spectral Analysis}

\ark564 is a bright, nearby ($z=0.024684$), narrow-line Seyfert 1 (NLS1)
galaxy, with luminosity of $L_{2-10~keV}=
(2.4-2.8)\times10^{43}~$ergs~s$^{-1}$ 
\citep[Paper~I]{Turner2001, Matsumoto2004}. In Paper-I we have 
discussed the \chandra observation and data
reduction of this source which we briefly summarize here. \ark564 was
observed with the \chandra High Energy Transmission Grating Spectrometer
(HETGS) in 2008 in three separate exposures totaling 250~ks.  We
followed the standard procedure to extract spectra. We used the software
package CIAO (Version 4.3) and calibration database CALDB (Version
4.4.2) developed by the \chandra X-ray center. We co-added the negative
and positive first-order spectra and built the effective area files
(ARFs) for each observation using the \emph{fullgarf} CIAO script. The
HETG-MEG spectra were analyzed using the CIAO fitting package
\emph{Sherpa}.

We co-added the spectra obtained with HETG-MEG and averaged the
associated ARFs. We fitted the continuum with absorbed power law plus
black body component and modeled all the statistically significant local
absorption ($z=0$) features with Gaussian components.  Further we used
the photoionization model fitting code PHotoionized Absorption Spectral
Engine \citep[PHASE;][]{Krongold2003}, to fit the typical warm absorber
features.  The best fit model of intrinsic absorption requires
two warm absorbers with two different ionization states
($\log U$\footnote{The ionization parameter $U=\frac{Q(H)}{4\pi r^{2}n_{H}c}$, 
where \emph{Q(H)} is the rate of H ionizing photons, \emph{r} is the distance 
to the source, \emph{$n_{H}$} is the hydrogen number density 
and \emph{c} is the speed of light. For the \ark564 spectrum, 
$\log U = \log \xi-1.85.$}$=0.38\pm0.02$ and $\log U=-1.3\pm0.13$), 
both with moderate outflow
velocities ($\sim 100~$km~s$^{-1}$) and relatively low line of sight
column densities ($\sim N_{H}=10^{20}~$cm$^{-2}$). For detailed spectral
analysis and best fit parameters, we refer readers to Paper-I.

Though most of the intrinsic absorption features in the source spectrum
are well fitted with two warm absorbers, there are residual absorption
line-like features in the spectral regions of $19.0-21.0~\AA$,
$17.0-17.5~\AA$ and $13.5-13.8~\AA$ each with individual significance of
$1.7-6.9$ sigma (Fig. 1). To check for the consistency of these
residuals and to confirm that these are not the artifacts of co-adding
spectra, we inspect individual \ark564 HETGS-MEG spectra from the 2008
observations.  After fitting the data with continuum, Galactic and
two-phase WAs model (Model-A) as noted above, the same residuals are
found in the same spectral region in all the three observations
(Fig. 2).  In the following sections, we discuss the identifications,
statistical significance and possible origin of these absorption lines.

\section{Discovery of high velocity outflows}

\subsection{Identification of the Absorption Lines}

The strongest residual features in the co-added spectrum are present at
$19.805\pm0.005~\AA$, $19.845\pm0.001~\AA$ and $20.250\pm0.011~\AA$
(observed frame, Fig. 1). Errors refer to $1~\sigma$ confidence level
throughout the paper, unless noted otherwise.  First we try to identify
the features at $19.805\pm0.005~\AA$ and $19.845\pm0.001~\AA$.  There is
no known instrumental feature near these energies (Chandra Proposers'
Observatory Guide, or POG). The z=0 lines are already included in the
model; there are no permitted lines with oscillator strength $>0.0001$
at wavelength of either $19.805, 19.845 $ or $20.250~\AA$.  At the
observed energies, there would be no intervening system with
$z_{WHIM}<z_{\ark564}$ from the warm-hot intergalactic medium 
\citep[WHIM;][]{Mathur2003, Nicastro2005}. Therefore, we assume 
that these absorption lines are intrinsic
to the source.  We identify these lines based on a combination of
chemical abundance and line strength (the oscillator strength $f>0.1$).
Given the detected wavelength and assuming a very broad range of
inflow/outflow velocities $-60,000$ to $60,000~$km~s$^{-1}$ the likely
candidates are \ovii $K\beta$ at $\lambda_{rest}=18.62\AA$, \oviii
$K\alpha$ at $\lambda_{rest}=18.96\AA$, \caxvii $k\beta$ at
$\lambda_{rest}=19.56\AA$, \caxvi $k\beta$ at
$\lambda_{rest}=20.617\AA$, \arxiv $k\beta$ at
$\lambda_{rest}=21.15\AA$, \caxvi $K\alpha$ at $\lambda_{rest}=21.45\AA$
and \ovii $k\alpha$ at $\lambda_{rest}=21.602\AA$. Considering that
argon and calcium are orders of magnitude less abundant than oxygen, and
because \ovii $K\alpha$ and \oviii $K\alpha$ are by far the strongest
possible lines, the most likely candidates are \oviii $K\alpha$ with
inflow velocities of $0.019c$ and $0.021c$, or \ovii $K\alpha$ with
outflow velocities of $0.105c$ and $0.103c$.

To distinguish between the two possibilities of inflow and outflow, we
search for possible associations of other lines such as \oviii $k\beta$,
\ovii $k\beta$, \ovi and/or \ov. We do not find any possible association
for inflows.  The absorption feature at $20.25\pm0.01\AA$ corresponds to
\ovi $k\alpha$ ($\lambda_{rest}=22.037\AA$) at the outflow velocity of
$0.103c$. We also found an absorption feature at $17.085\pm0.011~\AA$ with
$1.7\sigma$ significance, corresponding to \ovii $k\beta$ line at the
outflow velocity of $0.105c$.  The detection of \ovi $K\alpha$ and \ovii
$K\beta$ lines at the same velocity favors the outflow scenario.

\subsection{Statistical Significance of Absorption Lines}

To investigate whether the apparent absorption lines could be due to
statistical fluctuations, we calculate the probability of detection of
individual lines due to random statistical fluctuation. First we fit the
lines with negative gaussians of fixed width 0.001\AA, folding through
the detector response and leaving other parameters (centroid and
amplitude) free to vary. The addition of three gaussian lines at
$19.805\pm0.006~\AA$, $19.845\pm0.005~\AA$ and $20.250\pm0.005~\AA$ to
our previous model (Model A) improves the fit statistic by
$\Delta\chi^{2}=120, \Delta d.o.f.=6$, an improvement at more than
99.99\% confidence by the F-test (Bevington and Robinson 1992).  We
measured the equivalent width (EW) of lines at 
$19.805~\AA$, $19.845~\AA$ and $20.250\AA$
of $15.6\pm2.5~$m\AA, $16.5\pm2.4~$m\AA~ and $16.1\pm3.4~$m\AA~ 
respectively. Errors are at 1 sigma confidence level and are calculated
using the ``{\it projection}'' command in {\it Sherpa}. Thus the three
absorption lines are detected with $6.2\sigma$, $6.9\sigma$ and
$4.7\sigma$ significance respectively.  Further, using the Gaussian
probability distribution, we looked for the probability of detection of
these lines by chance. For the lines detected with $6.2\sigma$,
$6.9\sigma$ and $4.7\sigma$ significance, the probability of false
detection is $2.8\times10^{-10}$, $2.6\times10^{-12}$ and
$1.3\times10^{-6}$ respectively.  There are, however, 4801 wavelength
bins in our spectrum (these are all the wavelength bins in the
wavelength range of $1-25~\AA$, beyond which data is of poor
quality. The rest of the HETG spectrum was never used in any of the
analysis, even in Paper~I). Therefore the probability of finding
absorption lines at the observed significance anywhere in the spectrum
due to random statistical fluctuations is $0.0001$\%, $0.000001$\% and
$0.6$\% respectively.

For the outflow system with velocity of $0.105c$, we also detected the
\ovii K$\beta$ line with chance probability of $0.04$. Thus the combined
chance detection probability of detecting both \ovii K$\alpha$ and \ovii
K$\beta$ lines is $4 \times 10^{-6}$\%. Similarly for the other system
(at $0.103c$) the combined chance detection probability of detecting both
\ovii K$\alpha$ and \ovi K$\alpha$ is $4 \times 10^{-8}$\%. Thus we
conclude that the detected absorption lines are not due to random
statistical fluctuations, but are signatures of outflows.

\subsection{Other Absorption lines at blueshift of $\approx 0.1c$}

After confirming the high detection significance of \ovii and \ovi
absorption lines identified with high velocity outflows, we searched the
entire spectrum for other ionic transitions at similar blueshifts. We
detected two other absorption features at $17.351\pm0.009$ and
$13.625\pm0.011~\AA$ with EWs of $11.8\pm2.7$ and $5.0\pm1.7~$m$\AA$
(Fig. 3, Table 1).  The feature at $13.63\AA$ is also detected in the
HEG spectrum. The tentative identification of these features are \oviii
$K\alpha$ and \fexvii $K\alpha$ at outflow velocities $\sim 0.110 c$ and
$\sim 0.114 c$ respectively. The outflow velocities of these systems are
close to each other and to the above reported systems of high velocity
outflow.

\section{Photoionization Modeling of High Velocity Outflows}

To determine the physical properties of the absorber responsible for
producing highly blueshifted absorption features and to check for the
physical consistency of line identifications, we modeled the negative
Gaussian lines in the above fits with a photoionization model based on
the code PHASE (Krongold et al. 2003).  The PHASE code models
self-consistently more than 3000 X-ray bound-bound and bound-free
transitions imprinted by photoionized absorbers, given the ionization
state of the absorber, its column density, and its internal turbulent
motion. The parameters of the code are: 1) ionization parameter {\bf U};
2) equivalent hydrogen column density {\bf $N_{H}$}; 3) outflow velocity
of the absorbing material {\bf $V_{out}$}, and 4) micro-turbulent
velocity {\bf $V_{turb}$} of the material. The abundances have been set
at the Solar values \citep{Grevesse1993}. Usually, the micro-turbulent
velocity is not left free to vary, because different transitions due to
ionized gas are heavily blended, or because different velocity
components are also blended and cannot be resolved (e.g. Krongold et
al. 2003, 2005 for NGC 3783). In the case of \ark564, with the HETGS
resolution, it is possible to distinguish two velocity components (see
below).  So, despite the fact that the individual absorption lines
cannot be resolved, we have left the micro-turbulent velocity free to
vary.  We used the \ark564 spectral energy distribution (SED) from
Romano et al. (2004) to calculate the ionization balance of the
absorbing gas in PHASE, same as in Paper-I.

A PHASE component (which we call system 1) with ionization parameter of
$\log U=-0.60\pm0.38$ and column of $\log N_{H}=19.8\pm0.2~$cm$^{-2}$
successfully reproduced the \ovii K$\alpha$ and K$\beta$ lines at
$\sim19.805$ and $\sim17.08~\AA$, at the blueshift of
$32365\pm38~$km~s$^{-1}$ with respect to source (Fig. 4). The addition
of PHASE component ``System 1'' to Model-A significantly improves the
fit ($\Delta\chi^{2}=15, \Delta~d.o.f.=4$).  According to F-test, the
absorber is present at confidence level of $99.87$\%. A second low
ionization parameter component (we call this system 2) with $\log
U=-1.2\pm0.21$ and column of $\log N_{H}=20.0\pm0.3~$cm$^{-2}$ is also
required to fit other high velocity outflow absorption features such as
of \ovii and \ovin, with outflow velocity of $31735\pm59~$km~s$^{-1}$
(Fig. 4). Inclusion of this component improves the fit
($\Delta\chi^{2}=31$, $\Delta~d.o.f.=4$) over the previous model at a
confidence level of more than 99.999\%, according to F-test. The best
fit PHASE parameters are reported in Table 2 and the best fit model are
shown in Fig. 5. Successfully modeling the residuals with two PHASE
components robustly confirms the presence of high velocity outflows in
\ark564.

\section{Discussion}

\subsection{Comparison with Theoretical Models}

Several models suggest radiation pressure as the driving mechanism to
produce the typical low velocity outflows observed in Seyfert galaxies
\citep{Proga2002,Proga2004, Krolik1995, Dorodnitsyn2008}. But in
radiation driven disk-wind models outflow velocities depend on AGN
luminosity and these models cannot account for relativistic velocities
in AGNs with Seyfert-type luminosities \citep{Barai2011}. Relativistic
outflows in the UV have been detected only in most luminous broad
absorption line quasars. Indeed, \citet{Laor2002} and \citet{Ganguly2008}
have shown that the maximum outflow velocity is proportional to
$L^{0.6}$, close to what is expected from radiation pressure driven
winds.  Observations of relativistic outflows in X-rays in moderate
luminosity AGNs therefore pose intriguing puzzles; as shown in figure 6,
the relativistic outflows \citep[from][]{Tombesi2011} lie above the
Ganguly et al. line of maximum velocity. The relativistic outflow we
find in \ark564, shown as a $\star$ in figure 6, also lies above the
line. Thus these high velocity outflows appear not to be driven by
radiation line pressure mechanism.

\citet{King2010} shock wind models produce winds with velocities 
$v\sim~0.1c$, but in quasars accreting at Eddington limits. 
In this model a high velocity ionized outflow collides with the ISM 
of the host galaxy, losing much of its energy by efficient cooling 
resulting in a strongly shocked gas. 

In the multi-dimensional Monte Carlo simulations of AGN disk-wind models
of Sim et al., outflow velocities are {\it assumed} to be at escape
velocities, and not from ab-initio calculations \citep{Sim2008,
Sim2010}. It is possible that magnetic fields are important for
launching such winds, as in jets \citep{Sim2010}; observations of
relativistic outflows in moderate luminosity AGNs provide valuable
constraints to the theory of magneto-hydrodynamic winds.

The magneto-hydrodynamic accretion-disk wind models of
\citet{Fukumura2010a, Fukumura2010b} predict high-velocity ($v_{out}
\leq 0.6c$) outflows from ab-initio calculations. These models, however,
explain only the high-ionization high-velocity outflows, similar to
those observed by Tombesi et al. (2012). In these models, ultra-high
velocities are produced when UV to X-ray spectral slope is steep
($\alpha_{OX} \leq -2$), i.e. the AGNs are relatively UV bright (or
X-ray faint). Another feature of these models is that density scales as
$n(r) \propto r^{-1}$, leading to a clear prediction that the outflow
velocity depends on ionization parameter. Perhaps some of the model
assumptions can be modified to explain the ultra-high velocity warm
absorbers we present here.

\subsection{Mass and Energy Outflow Rates}

In Paper-I we have showed that the typical warm-absorbers in \ark564
($v_{out} \approx 100~$km~s$^{-1}$) are not energetic enough to provide
the effective feedback as required by theoretical models. Here we
estimate the total mass and kinetic energy outflow rate of relativistic
outflows in \ark564 to determine if these can possibly be important for
feedback.  To measure the mass and energy outflow rates, we must know
the location of the absorber which is measured only in a few sources
\citep[e.g. NGC~4051,][]{Krongold2007}. Since we do not know the
location of the high velocity outflow in \ark564, we estimate the mass
and energy outflow rates in several different ways.

It is often assumed that the observed outflow velocity is the escape
velocity at the launch radius $r$: i.e. $r
=\frac{2GM_{BH}}{v_{out}^{2}}$. There is no justification for this
assumption, as shown by \citet{Mathur2009}. Nonetheless, making this
assumption provides us with a lower limit on the absorber location.
\citet{Romano2004} determine the central black hole mass of \ark564 to
be $M_{BH}\leq~8\times 10^{6}M_{\odot}$. This leads to the minimum
distance of system~1 and system~2 absorbers of $r_{min}= 84r_{s}$ and
$r_{min}= 88r_{s}$ respectively (in units of the Schwarzschild radius;
for \ark564 $r_{s} = 7.8 \times 10^{-7}~$pc).  The estimate of maximum
distance from the central source can be derived assuming that the depth
$\Delta r$ of the absorber is much smaller than the radial distance of
the absorber ($\Delta r << r$) and using the definition of ionization
parameter ($U=\frac{Q(H)}{4\pi r^{2}n_{H}c}$), i.e.  $r \leq
r_{max}=\frac{Q(H)}{4 \pi U N_{H} c}$.  Using the best fit values of
ionization parameter and column density, we estimated the upper limits
on system~1 and system~2 absorber locations of $r_{max} = 5.4~$kpc and
$r_{max} = 13.6~$kpc respectively, which are not very interesting
limits.

For a bi-conical wind, the mass outflow rate is 
$\dot{M}_{out} \approx 1.2 \pi m_{p} N_{H} v_{out}r$ \citep{Krongold2007}. 
Substituting $r$ with $r_{min}$ and using outflow velocities of 
32365~km~s$^{-1}$ and 31735~km~s$^{-1}$, we obtain lower limit on mass 
outflow rates of $\dot{M}_{out} \geq 4.1 \times 10^{20}~$g~s$^{-1}$ 
and $\dot{M}_{out} \geq 4.2 \times 10^{20}~$g~s$^{-1}$ for 
system 1 and system 2 absorbers respectively. Similarly we obtained 
the constraints on kinetic luminosity of the outflows of 
$\dot{E}_{K}\geq 7.2 \times 10^{39}~$erg~s$^{-1}$ and 
$\dot{E}_{K}\geq 7.1 \times 10^{39}~$erg~s$^{-1}$ for the system 1 
and system 2 absorbers respectively. In comparison to the \ark564 
bolometric luminosity of $2.4 \times 10^{44}~$erg~s$^{-1}$, the total 
kinetic luminosity of these high velocity outflows is 
$\dot{E}_{K}/L_{bol}\geq 0.006$\%. This lower limit 
is much smaller than the ratio of $\dot{E}_{K}/L_{bol}\sim 0.5-5$\% expected 
if the outflow is to be important for feedback \citep{Hopkins2010, 
Silk1998,SO2004,Di2005}.However, with very large uncertainties between 
$r_{min}$ and $r_{max}$, the ratio between the mechanical power of these 
outflows and the bolometric luminosity cannot
be well constrained.

\subsection{Comparison with other high velocity outflows}

Before this work, the high velocity outflows were detected mostly with
\fexxv and/or \fexxvi absorption lines in hard X-ray band (\S 1). In a 
handful of quasar such as PG11211+143 \citep{Pounds2003a}, PG0844+349
\citep{Pounds2003b} and MR 2251-178 \citep{Gibson2005} high velocity
outflows were detected at soft X-ray energies, but these detections were
either based on absorption lines of low statistical significance or have
strong contamination from absorption lines from the halo of our Galaxy.

Here we present the detection of high velocity outflows in \ark564; this is
the first time such outflows are found in the soft X-ray band in a Seyfert
galaxy. We firmly establish the presence of high velocity WAs first
through identifying number of ionic transitions at similar velocity and
further successfully modeling these features with photoionization
models.  \citet{Papadakis2007} also detected an absorption line at 8.1
keV in the low resolution CCD spectra of \ark564 and assuming that this line
corresponds to \fexxvi, they suggested the presence of highly ionized,
absorbing material of $log N_{H} > 23~$cm$^{-2}$ outflowing with
relativistic velocity of 0.17c. If the presence of such a feature and
its identification are correct, then this is suggestive of a velocity
gradient with higher charge states such as \fexxvi at higher velocity.
Interestingly, the two ``other'' lines we identified in \ark564 (\S 3.0.3)
also have somewhat higher velocity than System 1 and System 2 and have
higher charge states. This is exactly as expected from the models of
\citet{King2010} and \citet{Fukumura2010a, Fukumura2010b}. 

Recently \citet{Tombesi2013} presented the connection between ultra-fast
outflows (UFOs) and soft X-ray WAs. They strongly suggest that these
absorbers represent parts of a single large-scale stratified outflow and
they continuously populate the whole parameter space (ionization,
column, velocity), with the WAs and the UFOs lying always at the two
ends of the distribution (Fig. 7).  The \ark564 low-velocity WAs
(Paper-I) and UFOs (Papadakis et al.) are in well agreement with linear
correlation fits from Tombesi et al. 2013.  However, our low-ionization
low-column high-velocity outflows in the \ark564 probe a completly
different parameter space. Figure~7 clearly shows that \ark564
high-velocity outflows have ionization parameter and column density as
of typical WAs, but much higher velocity, probing a distinct region in
velocity versus ionization/column parameter space.

\section{Conclusion}

We report on the discovery of high velocity outflows in the NLS1 Galaxy
\ark564. These absorbers are identified through multiple absorption
lines of \ovi, \ovii $K \alpha$, \ovii $K\beta$, \oviii and \fexvii at
blueshifts of $\sim 0.1c$ (with respect to the source) detected in the
\chandra HETG-MEG spectra. The two observed velocity components are well
fitted with two photoionization model components. Both absorbers have
low ionization parameter of $\log U =-0.60\pm0.38$ and $-1.2\pm0.21$
and low column densities of $\log N_{H} =19.8\pm0.2$ and
$20.0\pm0.3~$cm$^{-2}$ and are required at high significance of 99.87\%
and $>$ 99.99\% respectively.

Without knowing the location of the absorber, its mass and
energy outflow rates cannot be well constrained; we find 
$\dot{E}(outflow)/L_{bol}$ lower limit of 
$\geq 0.006$\% assuming a bi-conical wind geometry. Determining the absorber 
location is therefore very important for providing meaningful constraints. 
This can be achieved through studying the response of absorption 
lines to continuum
variations.  This is the first time that absorption lines with
ultra-high velocities are unambiguously detected in the soft X-ray
band. The presence of outflows with relativistic velocities in AGNs with
Seyfert-type luminosities is hard to understand and provides valuable
constraints to models of AGN outflows. Radiation pressure is unlikely to
be the driving mechanism for such outflows and magneto-hydrodynamic may be
involved.  Finding such relativistic outflows in several other AGNs and
measuring their mass/energy outflow rates is therefore important.

\noindent
{\bf Acknowledgement:} Support for this work was provided by the
National Aeronautics and Space Administration through Chandra Award
Number TM9-0010X issued by the Chandra X-ray Observatory Center, which
is operated by the Smithsonian Astrophysical Observatory for and on
behalf of the National Aeronautics Space Administration under contract
NAS8-03060. YK acknowledges support from CONACyT 168519 grant and 
UNAM-DGAPA PAPIIT IN103712 grant.

\clearpage

\bibliography{apj}

\clearpage
\begin{deluxetable}{lccccc}
\tabletypesize{\scriptsize}
\tablecaption{Absorption lines identified with relativistic 
outflows in the HETGS Spectrum of Ark 564.}
\tablewidth{0pt}
\tablehead{
\colhead{Ion} & \colhead{$\lambda_{obs}$} & \colhead{$\lambda_{rest}$} & 
\colhead{$v_{out}$} & \colhead{$EW$}\\
\colhead{} & \colhead{$\AA$} & \colhead{\AA} & \colhead{km~s$^{-1}$} & 
\colhead{m$\AA$}}
\startdata
\ovi              &  $20.250\pm0.005$  & $22.026$   &    $30834\pm67$   &  $16.1\pm3.4$  \\
\ovii $K\alpha$   &  $19.845\pm0.005$  & $21.602$   &    $31039\pm68$   &  $16.5\pm2.4$  \\
\ovii $K\alpha$   &  $19.805\pm0.006$  & $21.602$   &    $31581\pm81$   &  $15.6\pm2.5$  \\
\oviii            &  $17.351\pm0.009$  & $18.969$   &    $32199\pm139$   &  $11.8\pm2.7$  \\
\ovii $K\beta$    &  $17.085\pm0.011$  & $18.627$   &    $31463\pm173$   &  $4.0\pm2.4$  \\
\fexvii           &  $13.625\pm0.011$  & $15.015$   &    $34330\pm215$   &  $5.0\pm1.7$  \\
\enddata
\end{deluxetable}

\begin{table}[tp]
\begin{center}\centering
\caption{Model parameters for Relativistic Outflows}
\begin{tabular}{lcc}
\hline
Parameter & System 1 & System 2\\
\hline
$Log~U$  & $-0.60\pm0.38$ & $-1.2\pm0.21$ \\
$Log~N_{H}~($cm$^{-2})$ & $19.8\pm0.2$ &  $20.0\pm0.3$ \\
$V_{Turb}~($km~s$^{-1})$ & $85\pm14$ & $92\pm12$  \\
$V_{Out}~($km~s$^{-1})$ & $32365\pm38$ &  $31735\pm59$ \\
$\Delta\chi^{2}\tablenotemark{a}$  & 15   &  31   \\
$Log~\xi~($erg~s$^{-1}~cm)$  & $1.25\pm0.38$ & $0.65\pm0.21$ \\
\hline
\tablenotemark{a}{Improvement in $\chi^{2}$ to fit after adding model 
component}
\end{tabular}
\end{center}
\end{table}

\clearpage

\begin{figure}
\epsscale{.80}
\plotone{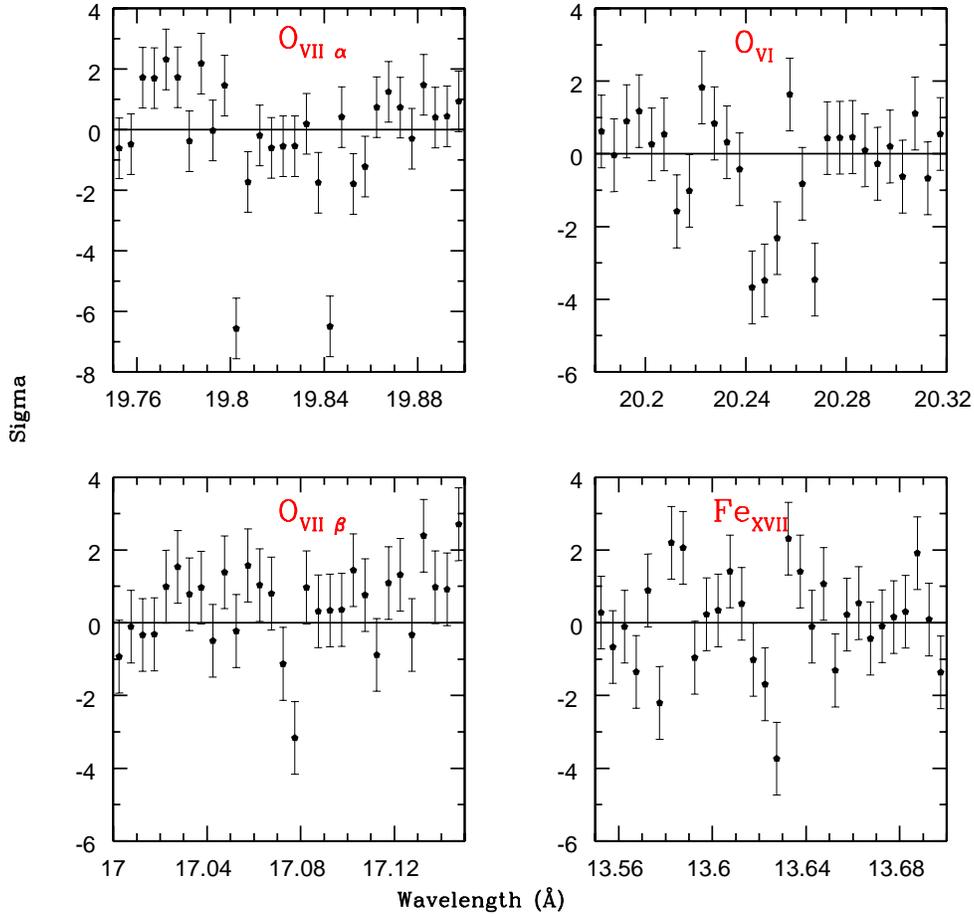}
\caption{Residuals of data:Model-A fit to the coadded spectrum of
Ark~564, in the observer frame (Model-A: Two component WA model plus
continuum, emission lines and local absorption, from Paper-I). The
possible transitions due to high velocity WAs are indicated with red
labels.  }.
\end{figure}

\clearpage

\begin{figure}
\epsscale{.80}
\plotone{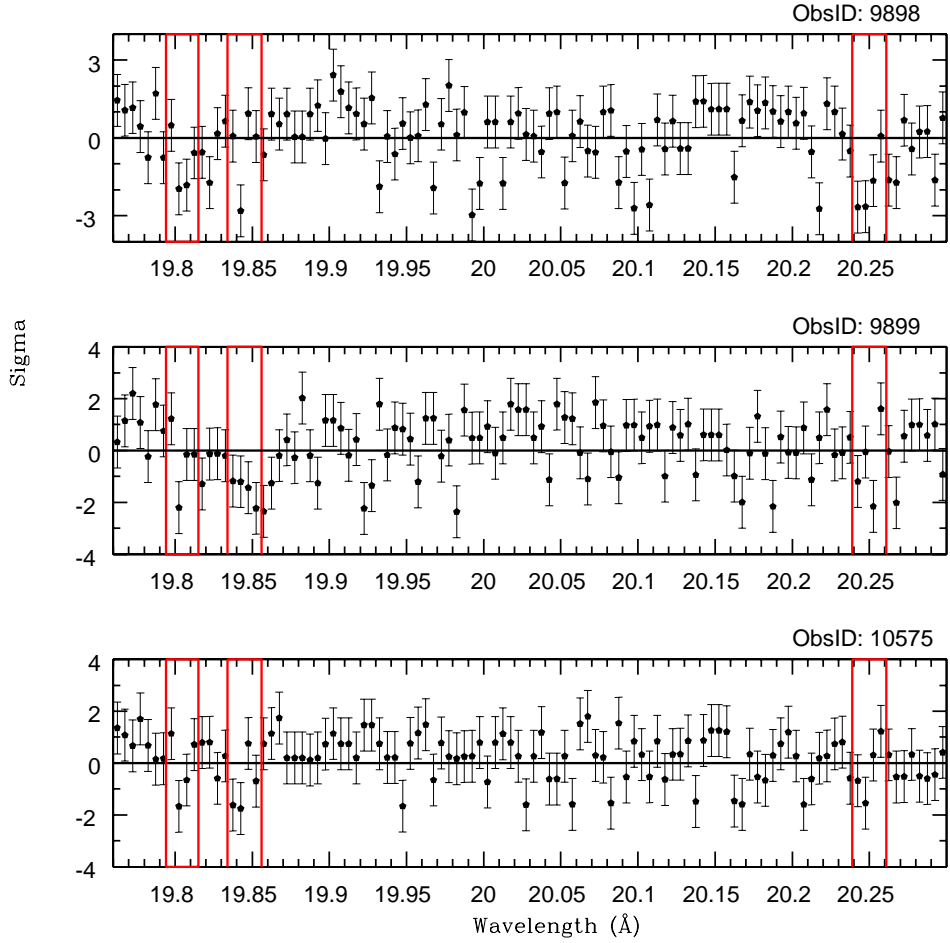}
\caption{Residuals of data:Model-A fit to individual spectra from the
2008 observations. The red boxes mark the regions near $19.804, 19.844$
and $20.24$\AA~ where the strongest lines are detected in the combined
spectrum. We see that the same lines are detected in the individual
spectra as well.}
\end{figure}

\clearpage

\begin{figure}
\epsscale{.80}
\plotone{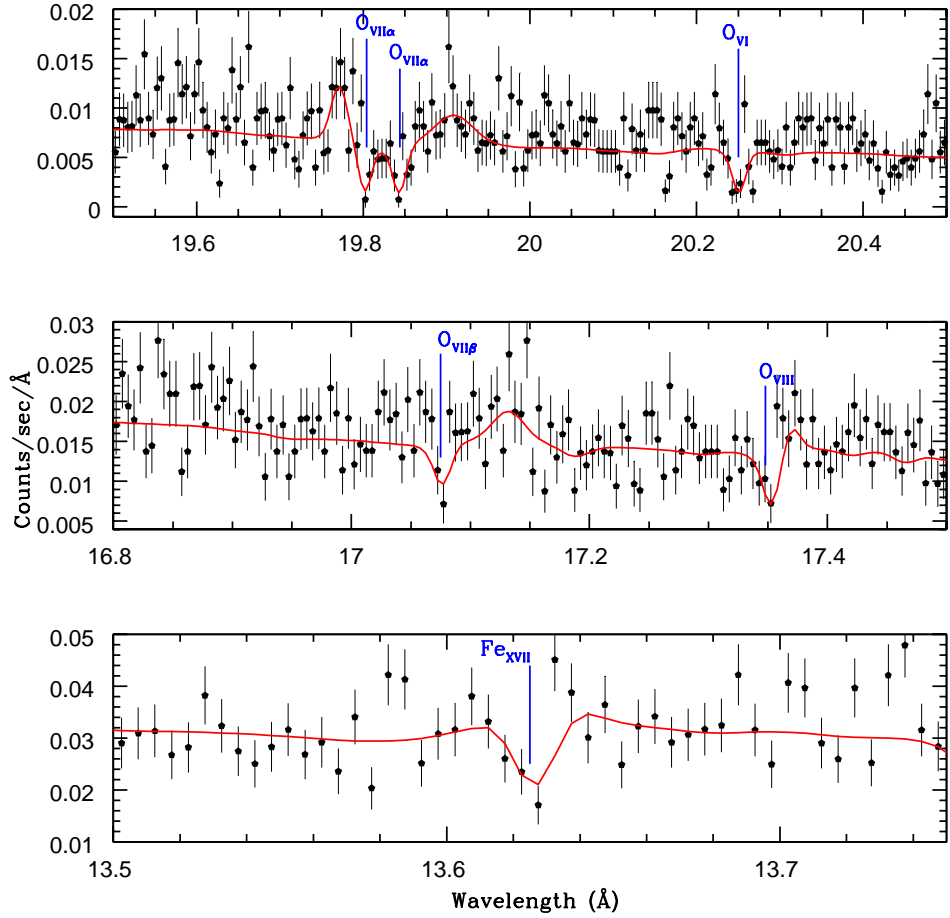}
\caption{The Ark~564 MEG spectrum fitted with Model-A plus six Gaussian
lines.  The line identifications of high-velocity outflow components are
indicated with blue labels.  The spectrum is presented in the observer
frame. }
\end{figure}

\clearpage
\begin{figure}
\epsscale{.80}
\plotone{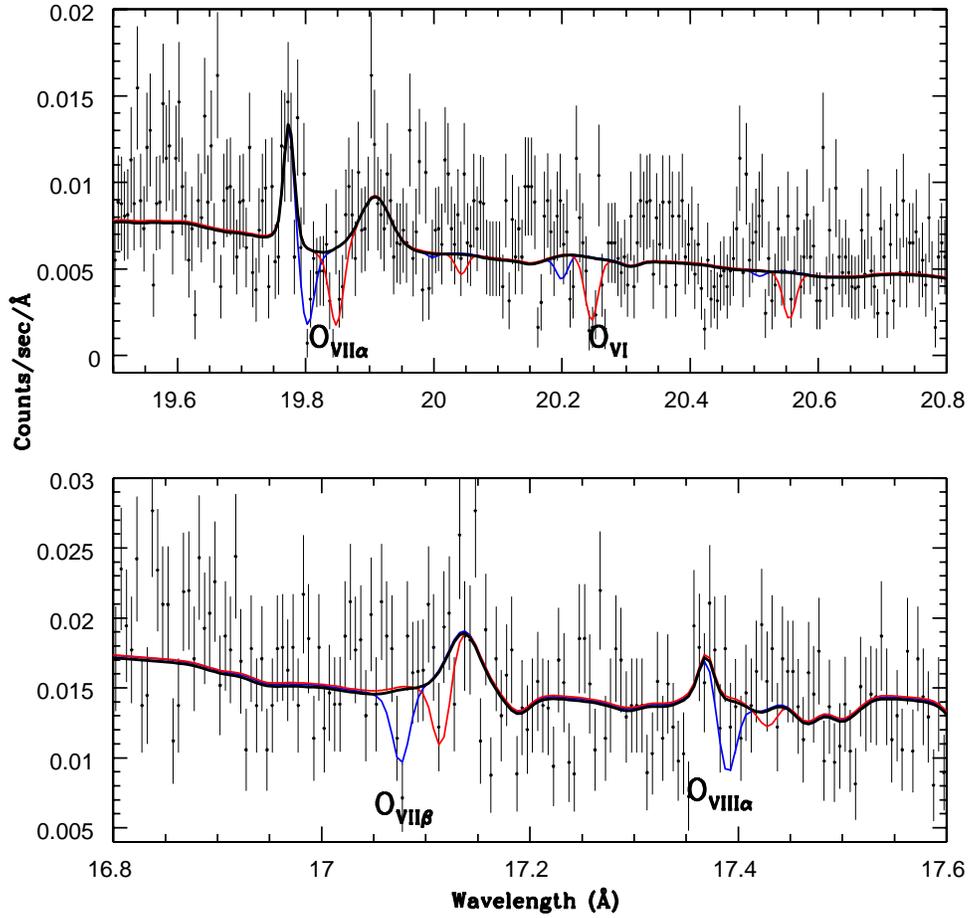}
\caption{The high velocity outflow absorption features fitted with two 
PHASE components of outflow velocity of 0.103c (red) and 0.105c (blue). 
The ionic transitions are labeled in black. 
The spectrum is presented in the observer frame.}
\end{figure}

\clearpage
\begin{figure}
\epsscale{.80}
\plotone{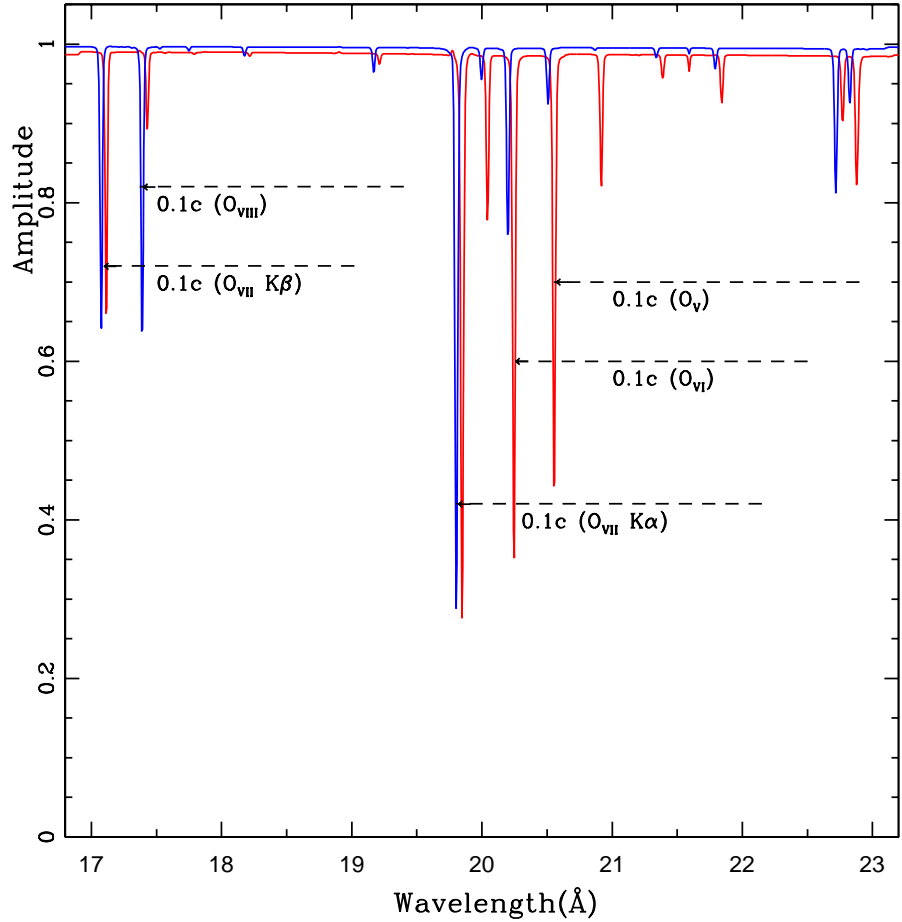}
\caption{The two component PHASE model, one with $v_{out}=0.105c$ (blue) 
and other with $v_{out}=0.103c$ (red), showing the enormous blueshift 
of absorption lines with respect to the source redshift.}. 
\end{figure}

\clearpage
\begin{figure}
\epsscale{.80}
\plotone{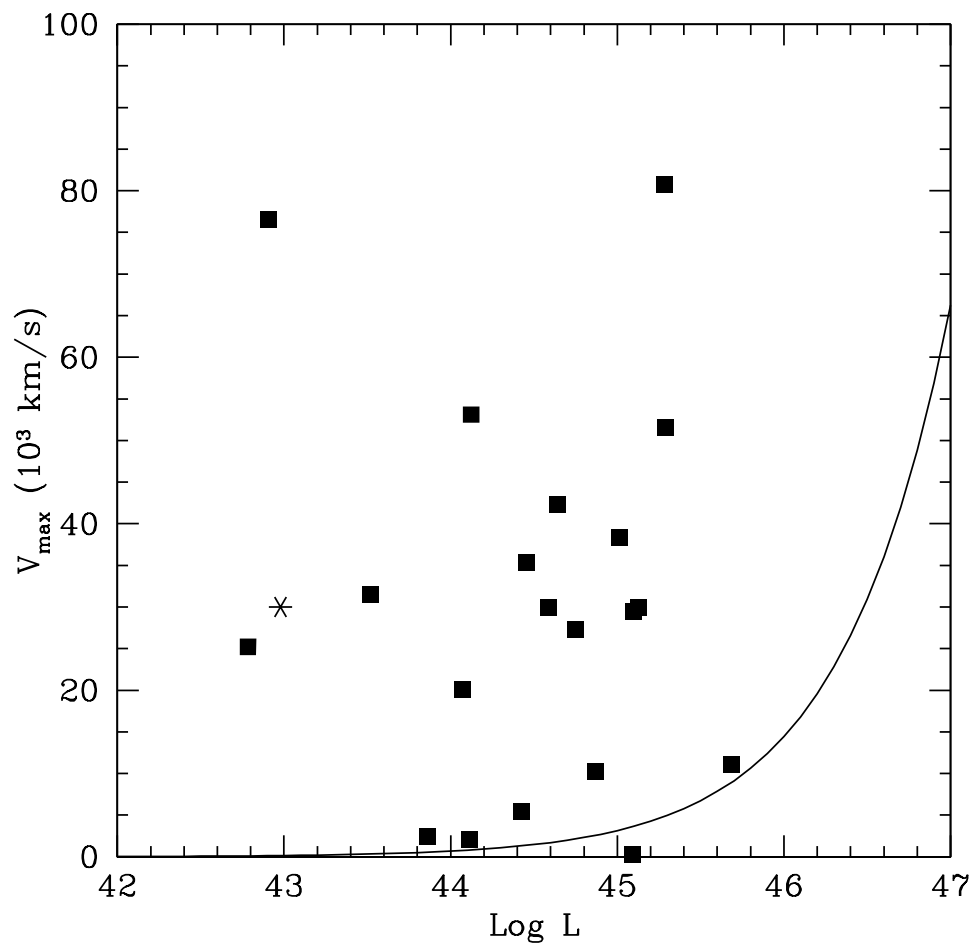}
\caption{Outflow  velocity plotted as a function of AGN luminosity. 
The solid line  represents the upper envelope relation from 
\citet{Ganguly2008} modified to plot the bolometric luminosity 
instead of the 3000\AA\  luminosity. The points are for the 
relativistic outflows in \citet{Tombesi2011}. Ark~564 ultra-fast 
outflow is shown by a star (this work). It is clear that these 
ultra-fast outflows are not confined  within the Ganguly et al. envelope.}
\end{figure}

\clearpage

\clearpage
\begin{figure}
\epsscale{.80}
\plotone{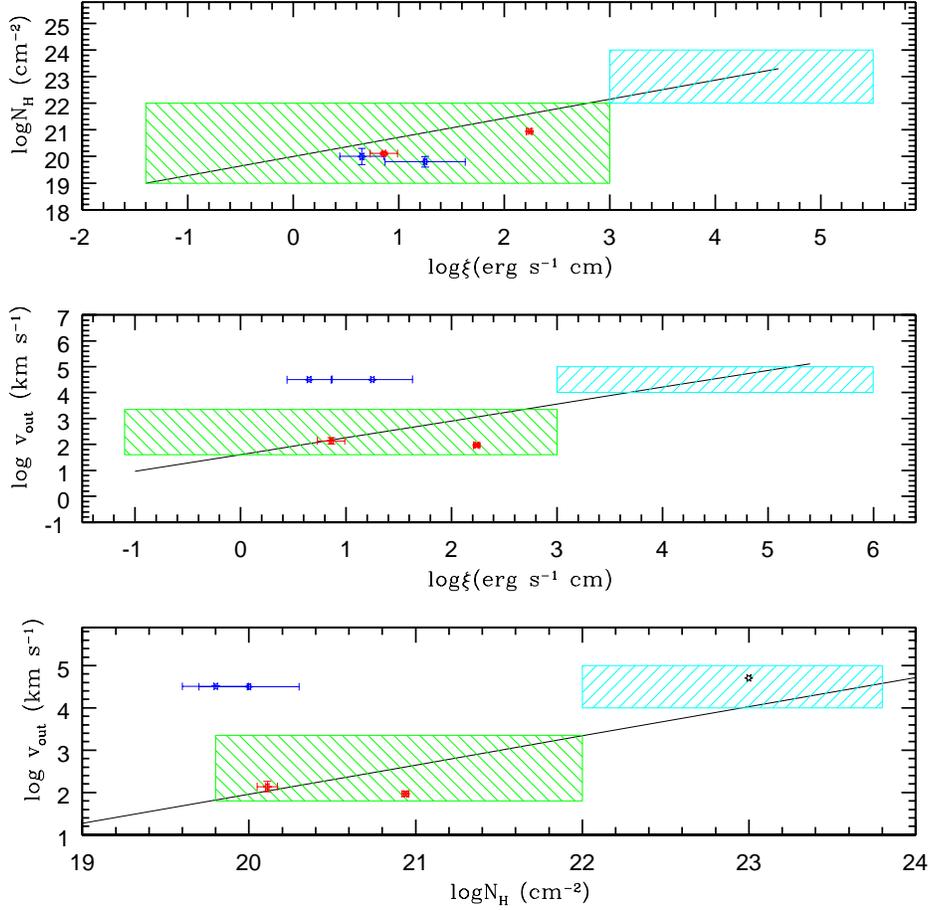}
\caption{The $\log \xi$ vs. $\log N_{H}$ (top panel), $\log \xi$ vs.
$\log v_{out}~$ (middle panel)and $\log N_{H}$ vs. $\log v_{out}~$
(bottom panel) for the low-velocity WAs (green striped region) and UFOs
(blue striped region) using data from \citet{Tombesi2013}. The solid
lines represent the correlation fits to low-velocity WAs and UFOs from
Tombesi et al. The data-points represent outflow parameters of Ark~564
low-velocity WAs (red; Paper-I), UFO (black; Papadakis et al. (2007))
and high-velocity (blue; this work).}
\end{figure}

\clearpage

\end{document}